\documentstyle[aps,twocolumn]{revtex}
\begin{document}
\noindent
{\large\bf{Comment on ``Composite Fermion Hofstadter Problem: Partially
Polarized Density Wave States in the $\nu=2/5$ ...''}}
\vspace*{0.8cm}

In a recent paper\cite{kukushkin}, magnetic field driven spin
transitions in fractional quantum Hall (FQH) states were
reported, and in particular, at $\nu=\frac23$ and
$\nu=\frac25$, weak features were observed at half polarization.
This would imply that the ground state energy is non-monotonic 
at half polarization. It was proposed in Ref.\cite{murthy} that
the system is then a charge density wave (CDW)
state of composite fermions with a subband
gap that could explain the observed plateaus at half 
polarization. Similar CDW state is also expected at $\nu=\frac23$
\cite{murthy}. Here we show that the proposed CDW state is not
the half polirized ground state at these two filling factors.
 
We compare the energies of the CDW state of 
composite fermions and the Halperin-(1,1,1) liquid state
\cite{book}. For $\nu =2/5$, where the composite fermions 
occupy the first two Landau levels,
the half-polarized state corresponds to a fully occupied 
$n=0$ $\uparrow$-spin Landau level and half-occupied $n=0$ 
$\downarrow$-spin and $n=1$ $\uparrow$-spin 
Landau levels. The CDW of Ref.\cite{murthy} is formed by $n=0$ 
$\downarrow$ and $n=1$ $\uparrow$ composite fermions
on a square lattice. We calculate the cohesive energy of this 
state from \cite{fukuyama}
\begin{equation}
E_{\rm CDW}=-\frac12 \sum_{\vec{Q},\sigma_1\sigma_2} 
    V_{\sigma_1\sigma_2}(\vec{Q})\Delta_{\sigma_1}(\vec{Q})
        \Delta_{\sigma_2}(-\vec{Q})
\end{equation}
where $\sigma=\uparrow$ and $\downarrow$, $V_{\downarrow\downarrow}$
is the Hartree-Fock potential for composite fermions on 
$n=0$ $\downarrow$-spin Landau level,  $V_{\downarrow\uparrow}$ 
is the Hartree potential for composite fermions on $n=0$ $\downarrow$-spin
and $n=1$ $\uparrow$-spin Landau levels and $V_{\uparrow\uparrow}$
is the Hartree-Fock potential for composite fermions on 
$n=1$ $\uparrow$-spin Landau level. The order parameter 
$\Delta _{\sigma }(\vec{Q})$ of the CDW corresponding to wave vector
$\vec{Q}$ for composite fermions with spin $\sigma$ is
taken to be non-zero only for 
reciprocal vectors: $\vec{Q} = (\pm Q_0,0), (0,\pm Q_0)$ and $(\pm Q_0,
\pm Q_0)$, $(\pm Q_0, \mp Q_0)$, where $Q^2_0 l^2=\pi $ ($l$ is the 
magnetic length for composite fermions). 
The energy of the Halperin-(1,1,1) state is calculated from
$E_{(1,1,1)}=-\frac1{4\pi}\int d^2 r V_{\rm eff}(r)[g(r)-1],$
where $g(r)=1-\exp(-r^2/2 l^2)$ is the correlation function for a fully 
occupied Landau level and 
$V_{\rm eff}(r) = \frac14\left[V^{\rm H}_{\downarrow\downarrow}(r) +  
      V^{\rm H}_{\uparrow\uparrow}(r) + 2V_{\uparrow\downarrow}(r)\right]$
[the subscript H refers to the Hartree term of the 
interactions in Eq. (1)], is the effective interaction between 
composite fermions.
 
The energy of the liquid state at half polarization is $-0.196
e^2/\varepsilon l$ and is {\it lower} than that of the CDW 
state, $-0.123 e^2/\varepsilon l$. However, it is not clear
that the Halperin-(1,1,1) state is indeed the ground state of the system. 
If we consider the Landau levels $n=0$ $\downarrow$ and $n=1$ 
$\uparrow$ as a two-level system and introduce pseudo-spin 
$\tau$ for the states at different levels, then Halperin-(1,1,1) 
state has $\tau=N/2$ and $\tau_z=0$, where $N$ is the 
number of composite fermions on $n=0$ $\downarrow$ and $n=1$ 
$\uparrow$ Landau levels. If this state is the correct 
ground state then the transition from a polarized to an 
unpolarized state of the system is just the rotation
of pseudo-spin vector from $\tau_z =-N/2$ to $\tau_z =N/2$ with 
fixed value of the total pseudo-spin $\tau=N/2$. But then
the ground state energy of the system is monotonic with
polarization of the system (quadratic function) without
any singularity at half polarization.

\begin{figure}
\begin{center}
\begin{picture}(100,130)
\put(0,0){\includegraphics{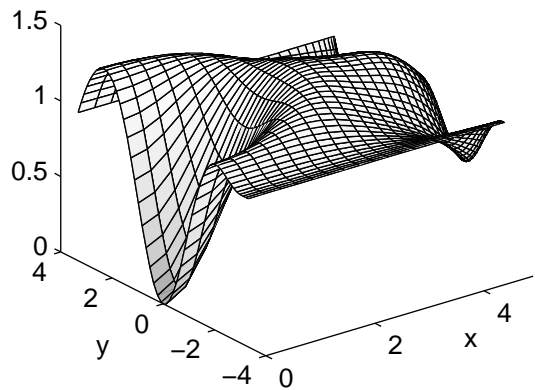}}
\end{picture}
\vspace*{1.8cm}
\caption{Pair-correlation function in spherical geometry.
Here, $x=2\sqrt{5.5}\sin(\phi/2)$, $y=2\sqrt{5.5}\sin(\theta/2-\pi/4)$.
}
  \end{center}
\end{figure}
 
We have also calculated the pair correlation functions 
in finite-size systems for original electrons. 
The pair correlation function for a 8-electron system at the
half-polarized state of $\nu =2/3$ in spherical geometry is 
smooth [Fig. 1] and do not exhibit any long-range order.
Based on these results we argue that the half-polarized 
ground state of two-dimensional electron system at filling factors 
$\nu=2/3$ and $\nu=2/5$ is not the CDW state proposed in 
Ref.\cite{murthy}. 

\vspace*{0.5cm}
\noindent
V.M. Apalkov$^1$, T. Chakraborty$^1$, P. Pietil\"ainen$^2$, and \hfil\break
K. Niemel\"a$^2$
\vspace*{0.5cm}

\noindent
$^1$Institute of Mathematical Sciences, Chennai 600 113, India,

\noindent
$^2$Theoretical Physics, University of Oulu,
Linnanmaa, FIN-90570 Oulu, Finland
\vspace*{0.5cm}



\begin{references}
\vspace*{-1.5cm}
\bibitem{kukushkin} I.V. Kukushkin, K. von Klitzing, and K. Eberl,
Phys. Rev. Lett. {\bf 82}, 3665 (1999).
\bibitem{murthy} G. Murthy, Phys. Rev. Lett. {\bf 84}, 350 (2000).
\bibitem{book} T. Chakraborty and P. Pietil\"ainen, {\it The
Quantum Hall Effects}, (Springer, New York, 1995), 2nd Edition.
\bibitem{fukuyama} D. Yoshioka and H. Fukuyama, J. Phys. Soc. Jpn. {\bf 47},
    394 (1979).
\end{references}
\end{document}